\documentclass[preprint,3p]{elsarticle}
\usepackage{amssymb}
\usepackage{amsthm}
\usepackage[mathlines]{lineno}
\usepackage{graphicx}
\usepackage{units}
\usepackage{url}
\usepackage{amsmath}
\usepackage{amsfonts}
\usepackage{bm}
\usepackage{textcomp}
\usepackage{subfigure}
\usepackage{multicol}
\usepackage{verbatim}
\usepackage{rotating}
\usepackage[colorlinks,linkcolor=red,citecolor=red]{hyperref}
\usepackage{float}
\usepackage{epsfig}
\usepackage{dcolumn}
\usepackage{bm}
\usepackage{color}
\usepackage{pstricks}
\usepackage{pst-node}
\usepackage{times}
\usepackage{indentfirst}
\usepackage[english]{babel}
\addto{\captionsenglish}{%

}
\journal{Physics Letters B}
\makeatletter

\newenvironment{figurehere}
  {\def\@captype{figure}}
  {}
\makeatother

\newcommand{\ssb}{\Sigma^0\bar{\Sigma}^0}
\newcommand{\ccb}{c\bar{c}}

\newcommand{\XXN}{\Xi^{0}\bar\Xi^{0}}
\newcommand{\XXB}{\Xi^{-}\bar\Xi^{+}}
\newcommand{\SSSN}{\Sigma(1385)^{0}\bar\Sigma(1385)^{0}}
\newcommand{\SSSM}{\Sigma(1385)^{-}\bar\Sigma(1385)^{+}}
\newcommand{\SSSP}{\Sigma(1385)^{+}\bar\Sigma(1385)^{-}}
\newcommand{\SSPM}{\Sigma(1385)^{\mp}\bar\Sigma(1385)^{\pm}}

\newcommand{\SX}{{\Sigma(1385)^{0}/\Xi^0}}
\newcommand{\SXB}{{\bar\Sigma(1385)^{0}/\bar\Xi^0}}

\newcommand{\EE}{e^+e^-}
\newcommand{\BB}{B\bar{B}}
\newcommand{\ppb}{p\bar{p}}
\newcommand{\psp}{\psi(3686)}
\newcommand{\jpsi}{J/\psi}
\newcommand{\ar}{\rightarrow}
\newcommand{\llb}{\Lambda\bar{\Lambda}}

\newcommand{\bfg}{\begin{figure}}
\newcommand{\efg}{\end{figure}}
\newcommand{\bitm}{\begin{itemize}}
\newcommand{\eitm}{\end{itemize}}
\newcommand{\bnum}{\begin{enumerate}}
\newcommand{\enum}{\end{enumerate}}
\newcommand{\btbl}{\begin{table*}}
\newcommand{\etbl}{\end{table*}}
\newcommand{\btbu}{\begin{tabular}}
\newcommand{\etbu}{\end{tabular}}
\newcommand{\bcl}{\begin{center}}
\newcommand{\ecl}{\end{center}}
\newcommand{\bbt}{\bibitem}
\newcommand{\beq}{\begin{equation}}
\newcommand{\eeq}{\end{equation}}
\newcommand{\beqr}{\begin{eqnarray}}
\newcommand{\eeqr}{\end{eqnarray}}

\begin{document}
\begin{frontmatter}
\title{{\bf \boldmath Study of $\jpsi$ and $\psp\ar\SSSN$ and $\XXN$}}
\author{
M.~Ablikim$^{1}$, M.~N.~Achasov$^{9,e}$, S. ~Ahmed$^{14}$, X.~C.~Ai$^{1}$, O.~Albayrak$^{5}$, M.~Albrecht$^{4}$, D.~J.~Ambrose$^{44}$, A.~Amoroso$^{49A,49C}$, F.~F.~An$^{1}$, Q.~An$^{46,a}$
% ...
\\
\vspace{0.2cm}
(BESIII Collaboration)\\
\vspace{0.2cm} {\it
$^{1}$ Institute of High Energy Physics, Beijing 100049, People's Republic of China\\
% ...
$^{4}$ Bochum Ruhr-University, D-44780 Bochum, Germany\\
$^{5}$ Carnegie Mellon University, Pittsburgh, Pennsylvania 15213, USA\\
% ...
$^{9}$ G.I. Budker Institute of Nuclear Physics SB RAS (BINP), Novosibirsk 630090, Russia\\
% ...
$^{14}$ Helmholtz Institute Mainz, Johann-Joachim-Becher-Weg 45, D-55099 Mainz, Germany\\
% ...
$^{44}$ University of Rochester, Rochester, New York 14627, USA\\
% ...
$^{46}$ University of Science and Technology of China, Hefei 230026, People's Republic of China\\
% ...
$^{49}$ (A)University of Turin, I-10125, Turin, Italy; (B)University of Eastern Piedmont, I-15121, Alessandria, Italy; (C)INFN, I-10125, Turin, Italy\\
% ...
$^{a}$ Also at State Key Laboratory of Particle Detection and Electronics, Beijing 100049, Hefei 230026, People's Republic of China\\
% ...
$^{e}$ Also at the Novosibirsk State University, Novosibirsk, 630090, Russia\\
% ...
}
}

\begin{abstract}
We study the decays of $\jpsi$ and $\psp$ to the final states $\SSSN$ and $\XXN$ based on a single baryon tag method
 using data samples of $(1310.6 \pm 7.0) \times 10^{6}$ $\jpsi$ and $(447.9 \pm 2.9) \times 10^{6}$ $\psp$ events collected with the BESIII detector at the BEPCII collider.
The decays to $\SSSN$ are observed for the first time. The measured branching fractions of $\jpsi$ and $\psp$ to $\XXN$ are in good agreement with, and much more precise than, the previously published results. The angular parameters for these decays are also measured for the first time. 
The measured angular decay parameter for $\jpsi\ar\SSSN$, $\alpha$ = $-0.64$ $\pm$ 0.03 $\pm$ 0.10, is found to be negative, different to the other decay processes in this measurement.
In addition, the \textquotedblleft 12\% rule" and isospin symmetry in the decays of $\jpsi$ and $\psp$ to $\Xi\bar\Xi$ and $\Sigma(1385)\bar{\Sigma}(1385)$ are tested.
\end{abstract}
\end{frontmatter}

\begin{multicols}{2}
\section{Introduction}
The decays of the charmonium resonances $\jpsi$ and $\psp$ [in the following, $\psi$ denotes both charmonium states $\jpsi$ and $\psp$] into baryon anti-baryon pairs ($\BB$) in $\EE$ annihilation have been extensively studied as a favorable test of perturbative quantum chromodynamics (QCD)~\cite{Farrar}.
These decays are assumed to proceed via the annihilation of the
constituent $\ccb$ pair into three gluons or a virtual photon.

It is interesting that the $\psp$ decay to a specific final state is strongly suppressed relative to the same final state in $\jpsi$ decay according to the annihilation decay of heavy quarkonium. The ratio of branching fractions for $\psi$ decaying into the same final states is predicted from factorization~\cite{Appelquist:1974zd} to be
$
\frac{{\cal B}(\psp\ar X)}{{\cal B}(J/\psi\ar X)}
 \approx 12\%$,
where $X$ denotes any exclusive hadronic decay mode or the $\ell^{+}\ell^{-} (\ell=e, \mu)$ final state.  This expectation is usually called the \textquotedblleft 12\% rule". This rule was first observed to be violated in the decay of $\psi$ into the final state $\rho\pi$.   A broad variety of reviews
of the relevant theoretical and experimental results~\cite{yfgu} conclude that the current theoretical explanations are unsatisfactory. Although the branching fractions for $\psi$ decays into baryon pairs have been measured extensively~\cite{PDG2012},
uncertainties are still large for many decays; \emph{e.g.} the world average values of the branching fractions for $\jpsi$ and $\psp\to \XXN$ are $(1.20 \pm 0.24)\times 10^{-3}$ and $(2.07 \pm 0.23)\times10^{-4}$~\cite{PDG2012}, respectively. In particular, $\psi\ar\SSSN$ has not yet been observed.

By hadron helicity conservation, the angular distribution of the process $\EE\ar\psi\ar\BB$ is expressed as
\begin{equation}\label{alpha}
\frac{dN}{d\cos\theta}\propto1+\alpha\cos^{2}\theta,
\end{equation}
where $\theta$ is the angle between the baryon and the beam directions in the $\EE$ center-of-mass (CM) system and $\alpha$ is a constant, which has widely been investigated in theory and experiment~\cite{Franklin:1983ve}.
Theoretically, the value of $\alpha$ is discussed in the framework of many models, such as quark mass effects~\cite{ppbref02}, or electromagnetic effects~\cite{ppbref01}, which generally predict $0<\alpha< 1$. 
BES measured the angular distribution of $\jpsi\ar\Sigma^0\bar\Sigma^0$ and obtained a negative 
$\alpha$ with poor precision~\cite{angularSig}. H.~Chen {\it et al.}~\cite{Chen:2006yn} explained
that the angular distribution for $\psi\ar\BB$ could be negative when
rescattering effects of baryon and anti-baryon in heavy quarkonium decays are taken into consideration.
Thus, experimental measurements of $\alpha$  are helpful to test the helicity conservation rule and the validity of the various theoretical approaches.
In previous experiments, the angular distributions for charmonium decays to baryon pairs,
such as $\psi\ar\ppb, \llb, \ssb, \XXB$, and $\SSPM$~\cite{ppbref, Ablikim:2006aw, Jximxip04, Ablikim:2016iym}, were measured.
However, angular distributions for the decays $\psi\ar\SSSN$ and $\XXN$ have not yet been measured.

In this Letter, we report the most precise measurements of the branching fractions and angular distributions for $\psi\ar\SSSN$ and $\XXN$ based on the data samples of $(1310.6 \pm 7.0) \times 10^{6}$ $J/\psi$~\cite{Jpsi} and $(447.9 \pm 2.9) \times 10^{6}$ $\psp$~\cite{Psip09, Psip} events collected with the BESIII detector at BEPCII.
\section{BESIII detector and Monte Carlo simulation}
\label{sec:detector}
BEPCII is a double-ring $\EE$ collider that has reached a peak luminosity of $10^{33}~\rm{cm}^{-2}\rm{s}^{-1}$
at a CM energy of 3.773 GeV.
The cylindrical core of the BESIII detector consists of a helium-based main drift
chamber (MDC), a plastic scintillator time-of-flight (TOF) system, and
a CsI(Tl) electromagnetic calorimeter (EMC), which are all enclosed in
a superconducting solenoidal magnet with a field strength 
of 1.0 T for the $\psp$ data and $\jpsi$ data taken in 2009, and 0.9 T for the $\jpsi$ data taken in 2012.
The solenoid is supported by an octagonal flux-return yoke with
resistive plate counter modules interleaved with
steel as muon identifier. The acceptance for charged particles and photons is 93\% of
the $4\pi$ stereo angle, and the charged-particle momentum resolution at 1 GeV/$c$ is 0.5\%.
The photon energy resolution is 2.5\% (5\%) at 1.0 GeV in the barrel region (end caps regions).
More details about the experimental apparatus can be found in Ref.~\cite{BESIII}.

The response of the BESIII detector is modeled with Monte Carlo (MC) simulations
using a framework based on \textsc{geant}{\footnotesize 4}~\cite{geant4}.
The production of $\psi$ resonances is simulated with the \textsc{kkmc} generator~\cite{kkmc},
the subsequent decays are processed via
\textsc{evtgen}~\cite{evt2} according to the measured branching fractions
provided by the Particle Data Group (PDG)~\cite{PDG2012}, and the
remaining unmeasured decay modes are generated with
\textsc{lundcharm}~\cite{lund}. To determine the detection efficiencies for $\psi\ar\SSSN$ and $\XXN$,
one million MC events are generated for each mode taking into account for the angular distribution 
 with $\alpha$ value measured in this analysis.
 The decays of the baryons $\Sigma(1385)^{0}$, $\Xi^{0}$, and $\Lambda$ in the signal channels are simulated exclusively, taking into account the angular distributions via \textsc{evtgen}~\cite{evt2},
 while the anti-baryons are set to decay inclusively.

\section{Event selection}
\label{sec:evt_sel}
The selection of $\psi\ar\SSSN$ and $\XXN$ events via a full reconstruction of both $\SX$ and $\SXB$ suffers from low reconstruction efficiency and large systematic uncertainty.

To achieve higher efficiency and reduce the systematic uncertainty, a single baryon $\SX$ tag technique is employed, without including the anti-baryon mode tag due to the imperfection of the simulation related to the effect of annihilation for anti-proton.  The $\SX$ is reconstructed in its decay to $\pi^{0}\Lambda$ with the subsequent decays $\Lambda\ar p\pi^{-}$ and $\pi^{0}\ar\gamma\gamma$. 
The charged tracks are required to be reconstructed in the MDC with good helix fits and within the angular coverage of the MDC ($|\cos\theta|<0.93$, where $\theta$ is the polar angle with respect to the $e^{+}$ beam direction).
Information from the specific energy loss measured in the MDC ($dE/dx$) and from the TOF are combined to form particle identification (PID) confidence levels for the hypotheses of a pion, kaon, and proton.
Each track is assigned to the particle type with the highest confidence level. At least one negatively charged pion and one proton are required. 
Photons are reconstructed from isolated showers in the EMC. The energy deposited
in the nearby TOF counter is included to improve the reconstruction
efficiency and energy resolution. Photon energies are required to be
greater than 25 MeV in the EMC barrel region ($|\cos\theta|<0.8$) or
greater than 50 MeV in the EMC end cap
($0.86 < |\cos\theta| < 0.92$). The showers in the angular range between
the barrel and the end cap are poorly reconstructed and are excluded from
the analysis. Furthermore, the EMC timing of the photon candidate must be
in coincidence with collision events, $0\leq t\leq 700$ ns, to
suppress electronic noise and energy deposits unrelated to the collision events. At least two good photon candidates are required. 

In order to reconstruct the $\pi^{0}$ candidates, a one-constraint (1C) kinematic fit is employed for all $\gamma\gamma$ combinations, constraining the invariant mass of two photons to the $\pi^0$ nominal mass, combined with the requirement of $|\Delta E|/P_{\pi^0} < 0.95$, where $\Delta E$ is the energy difference between the two photons and $P_{\pi^{0}}$ is the $\pi^{0}$ momentum,
and the $\chi^{2}_{1C} < 20$ to suppress non-$\pi^0$ backgrounds.

To reconstruct the $\Lambda$ candidates,
a vertex fit is applied to all $p\pi^{-}$ combinations; the ones characterized by $\chi^{2} < 500$ are kept for further analysis. The $p\pi$ invariant mass is required to be within
5 MeV/$c^{2}$ of the nominal $\Lambda$ mass, determined by optimizing the
figure of merit FOM = $\frac{S}{\sqrt{S + B}}$, where $S$ is the
number of signal events and $B$ is the number of background events based on the MC simulation. To further suppress the background, the decay length of $\Lambda$ is required to be larger than zero.
The $\SX$ candidates are reconstructed with $\Lambda$ and $\pi^0$ candidates by minimizing the variable $|M_{\pi^{0}\Lambda}-M_{\SX}|$, where $M_{\pi^0\Lambda}$ is the invariant mass of the $\pi^0\Lambda$ pair, and $M_{\SX}$ is the nominal mass of $\SX$.

The anti-baryon candidate $\SXB$  is inferred by the mass recoiling against the selected $\pi^{0}\Lambda$ system,
\begin{equation}
M^{\rm recoil}_{\pi^{0}\Lambda} = \sqrt{(E_{\rm CM}-E_{\pi^{0}\Lambda})^{2} - \vec{p}^{2}_{\pi^{0}\Lambda}},
\end{equation}
where $E_{\pi^{0}\Lambda}$ and $\vec{p}_{\pi^{0}\Lambda}$ are the energy and momentum of the selected 
$\pi^{0}\Lambda$ system, and $E_{\rm CM}$ is CM energy.
Figure~\ref{scatterplot} shows the scatter plot of $M_{\pi^0\Lambda}$ versus $M^{\rm recoil}_{\pi^0\Lambda}$.  Clear accumulations of events corresponding to the signals of $\psi\ar\SSSN$ and $\XXN$ decays are observed.
The distributions of $M_{\pi^0\Lambda}$ with the additional requirement of the $M^{\rm recoil}_{\pi^0\Lambda}$ within $\pm80$ MeV/$c^{2}$ around $M_{\Sigma(1385)^0}$ or $\pm50$ MeV/$c^{2}$ around $M_{\Xi^0}$ are shown in Fig.~\ref{mass_pi0lambda}. Clear $\SX$ signals are observed.
\begin{figurehere}
\bcl
\subfigure{\includegraphics[width=0.38\textwidth]{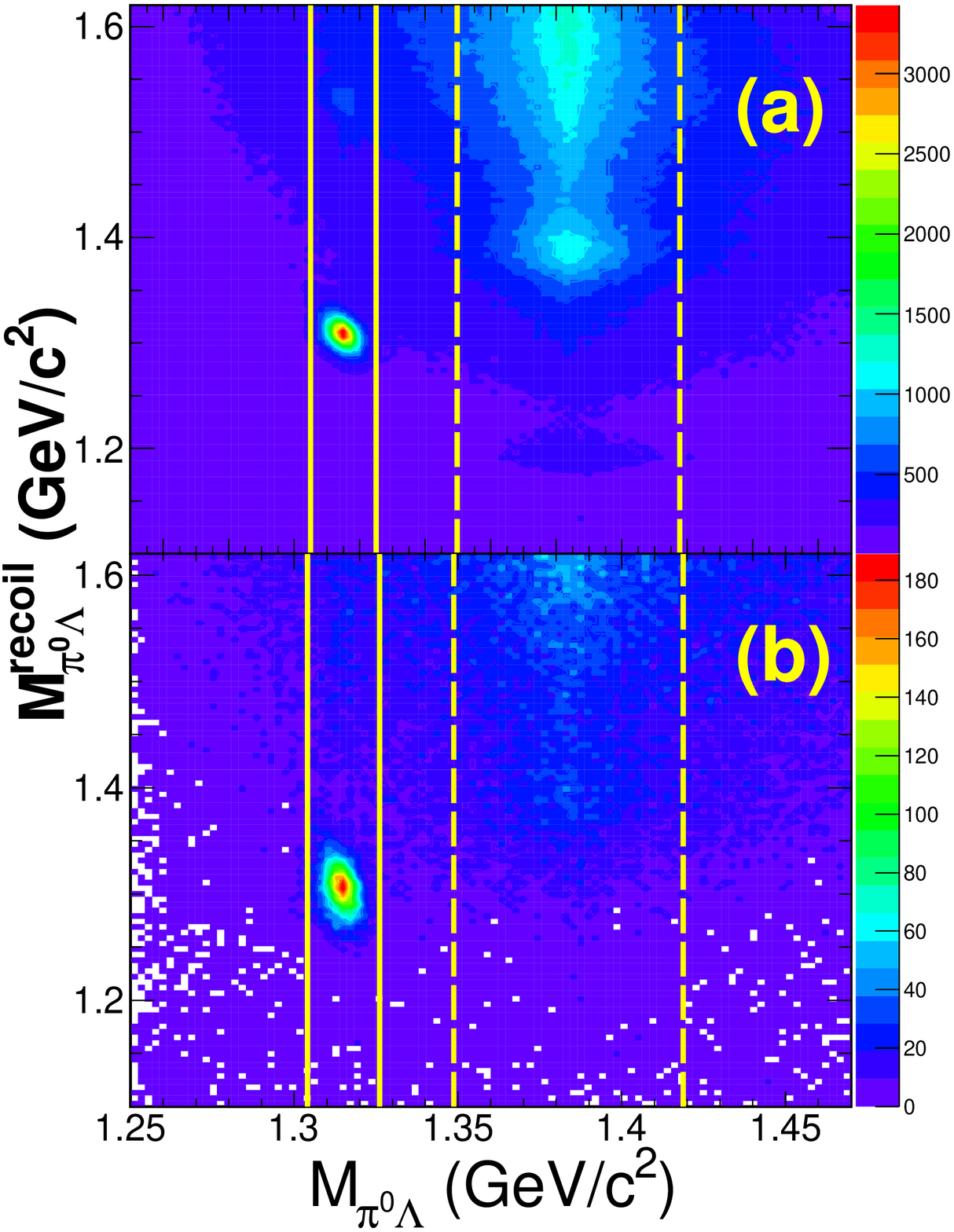}}
\caption{Scatter plots of $M_{\pi^{0}\Lambda}$ versus $M^{\rm recoil}_{\pi^{0}\Lambda}$ for (a) $\jpsi$ and (b) $\psp$ data. The dashed lines denote the $\Sigma(1385)^0$ signal region,
and the solid lines denote the $\Xi^0$ signal region.}
\label{scatterplot}
\ecl
\end{figurehere}
\begin{figurehere}
\bcl
\subfigure{\includegraphics[width=0.38\textwidth]{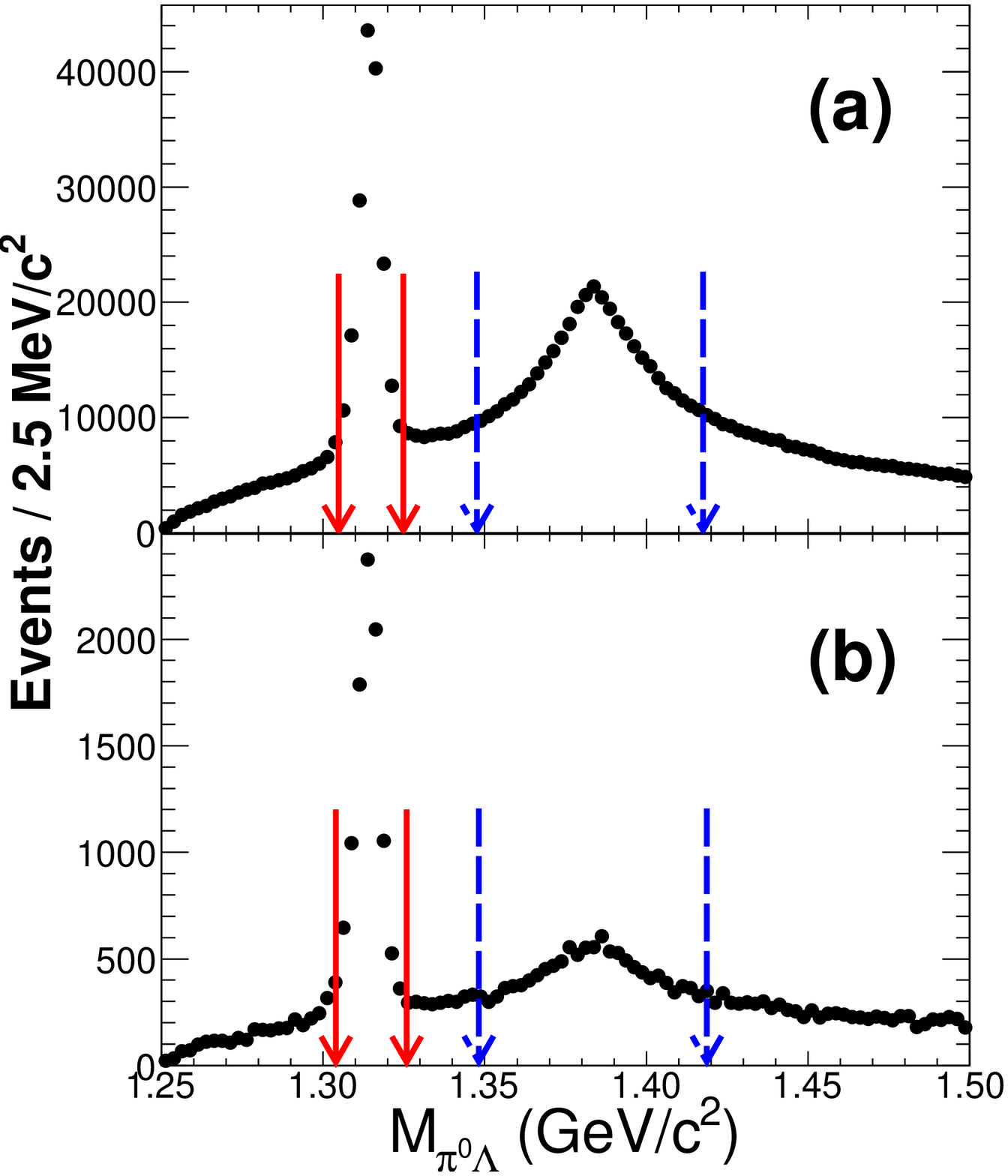}}
\caption{Distribution of $M_{\pi^{0}\Lambda}$ for (a) $\jpsi$ and (b) $\psp$ data. The arrows denote the applied requirements, where  the dashed arrows the $\Sigma(1385)^0$ signal region and the solid arrows show the $\Xi^0$ signal region.}
\label{mass_pi0lambda}
\ecl
\end{figurehere}

To determine signal yields, the mass of $\pi^{0}\Lambda$ is required to be within $\pm$34 MeV/$c^{2}$ for $\jpsi\ar\SSSN$, $\pm$10 MeV/$c^{2}$ for  $\jpsi\ar\XXN$, $\pm$35 MeV/$c^{2}$ for $\psp\ar\SSSN$, and $\pm$11 MeV/$c^{2}$ for $\psp\ar\XXN$, around the nominal mass of $\Sigma(1385)^{0}$/$\Xi^{0}$; the requirements are optimized by the FOM. For the $\psp$ decays, the requirements of $|M^{\rm recoil}_{\pi^{+}\pi^{-}}-M_{\jpsi}| > 0.005$ GeV/$c^{2}$ and
$|M^{\rm recoil}_{\pi^{0}\pi^{0}}- M_{\jpsi}| > 0.015$ GeV/$c^{2}$
 are used to suppress the backgrounds $\psp\ar\pi\pi\jpsi$, 
 where $M^{\rm recoil}_{\pi^+\pi^-}$ and $M^{\rm recoil}_{\pi^0\pi^0}$
 are the recoil masses of any $\pi^+\pi^-$ and $\pi^0\pi^0$ combination if found, and $M_{\jpsi}$ is the $\jpsi$ nominal mass according to the PDG~\cite{PDG2012}.
\section{Background study}
\label{sec:background}
The data collected at CM energies of 3.08 GeV (30 pb$^{-1}$)~\cite{Jpsi} and 3.65 GeV (44 pb$^{-1}$)~\cite{Psip} are used to estimate the
contributions from the continuum processes $\EE\ar\SSSN$ and $\XXN$. By applying the same event selection criteria,
only a few events survive and do not form any obvious peaking structures in the $\SXB$ signal regions in the corresponding $M^{\rm recoil}_{\pi^{0}\Lambda}$ distributions. Taking into account the normalization of the luminosity and CM energy dependence of the cross section, the QED backgrounds are found to be negligible.

The contamination from other background sources is analyzed using samples of MC simulated events of generic $\psi$ decays that contain the same number of events as the data. After applying the same event selection, it is found that the peaking backgrounds  for the $\psi\ar\SSSN$ mode mainly come from
$\psi\ar\XXN$, $\XXB$, $\Sigma(1385)^{-}\bar\Sigma(1385)^{+}$, $\Xi(1530)\bar\Xi$ + c.c., and $\pi^{0}\Lambda\bar\Sigma(1385)^{0}$, where the branching fractions for $\psi\ar\Xi(1530)\bar\Xi$ + c.c. and $\pi^{0}\Lambda\bar\Sigma(1385)^{0}$ are taken from the isospin partner modes $\jpsi\ar\Xi(1530)\bar\Xi$ + c.c.\cite{PDG2012} and $\pi^{-}\Lambda\bar\Sigma(1385)^{+}$\cite{Ablikim:2016iym} based on the assumption of 12\% rule.
For the $\jpsi\ar\XXN$ mode, the peaking backgrounds are found to be from $\jpsi\ar\XXB$, $\gamma\eta_{c}(\gamma\ssb, \gamma\XXN)$, $\Sigma^{0}\bar\Sigma(1385)^{0}$, and $\ssb$.
For the $\psp\ar\XXN$ mode, the peaking background is from $\psp\ar\ssb$, and other backgrounds are found to be distributed smoothly in $M^{\rm recoil}_{\pi^0\Lambda}$ mass spectrum.

The final states of baryon and anti-baryon decays both include a neutral pion with almost the same momenta. 
The $\pi^0$ from the anti-baryon can be wrongly combined with the $\Lambda$ in the $\SX$ reconstruction.
As a result, the wrong combination background (WCB) in the $\pi^{0}\Lambda$ mass spectrum is inevitable. This background is studied by the MC simulation.

\section{Results}
\subsection{Branching fraction}
\label{sec:branching}
\begin{figure*}[!hbt]
\bcl
\subfigure{\includegraphics[width=0.45\textwidth]{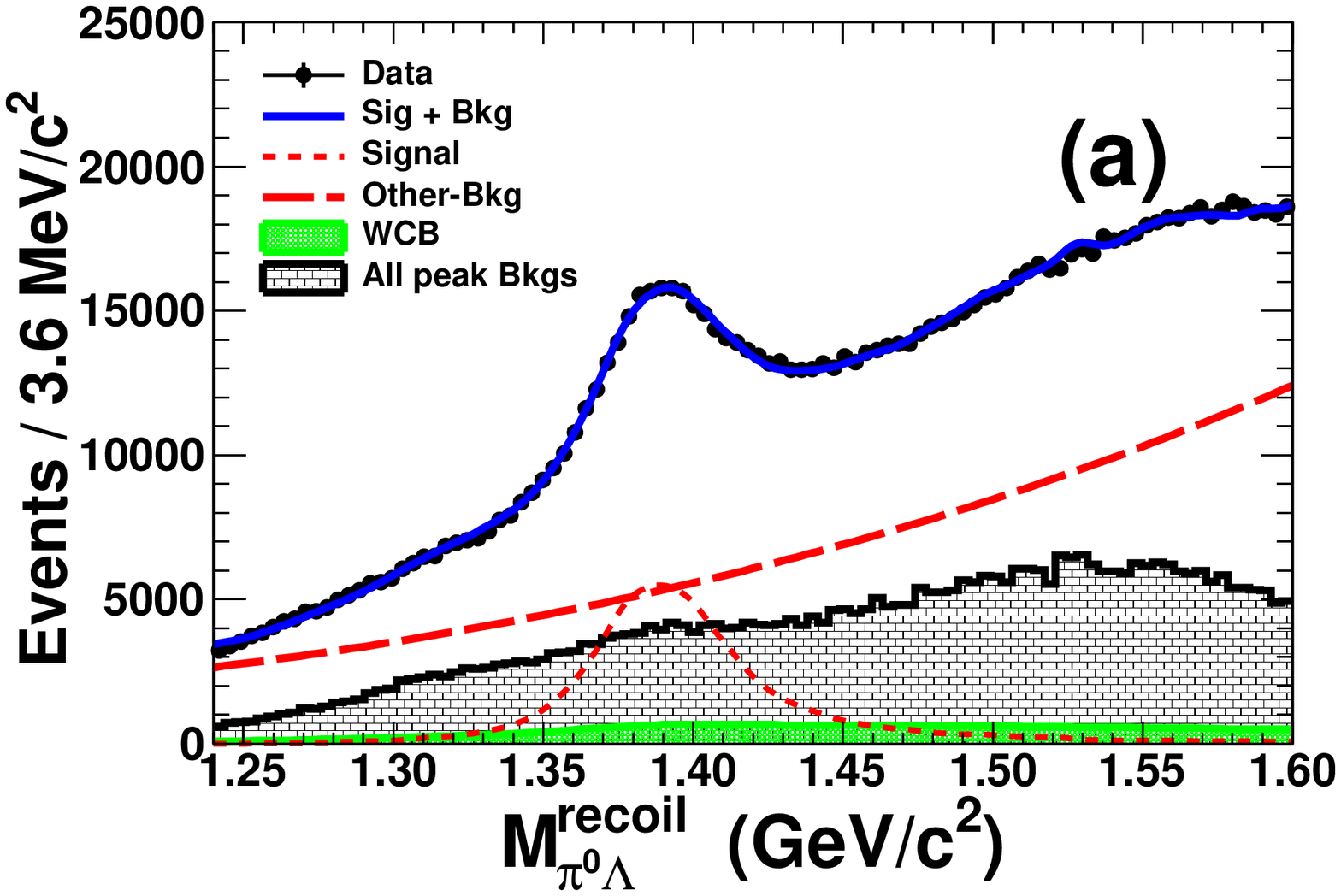}}
\subfigure{\includegraphics[width=0.45\textwidth]{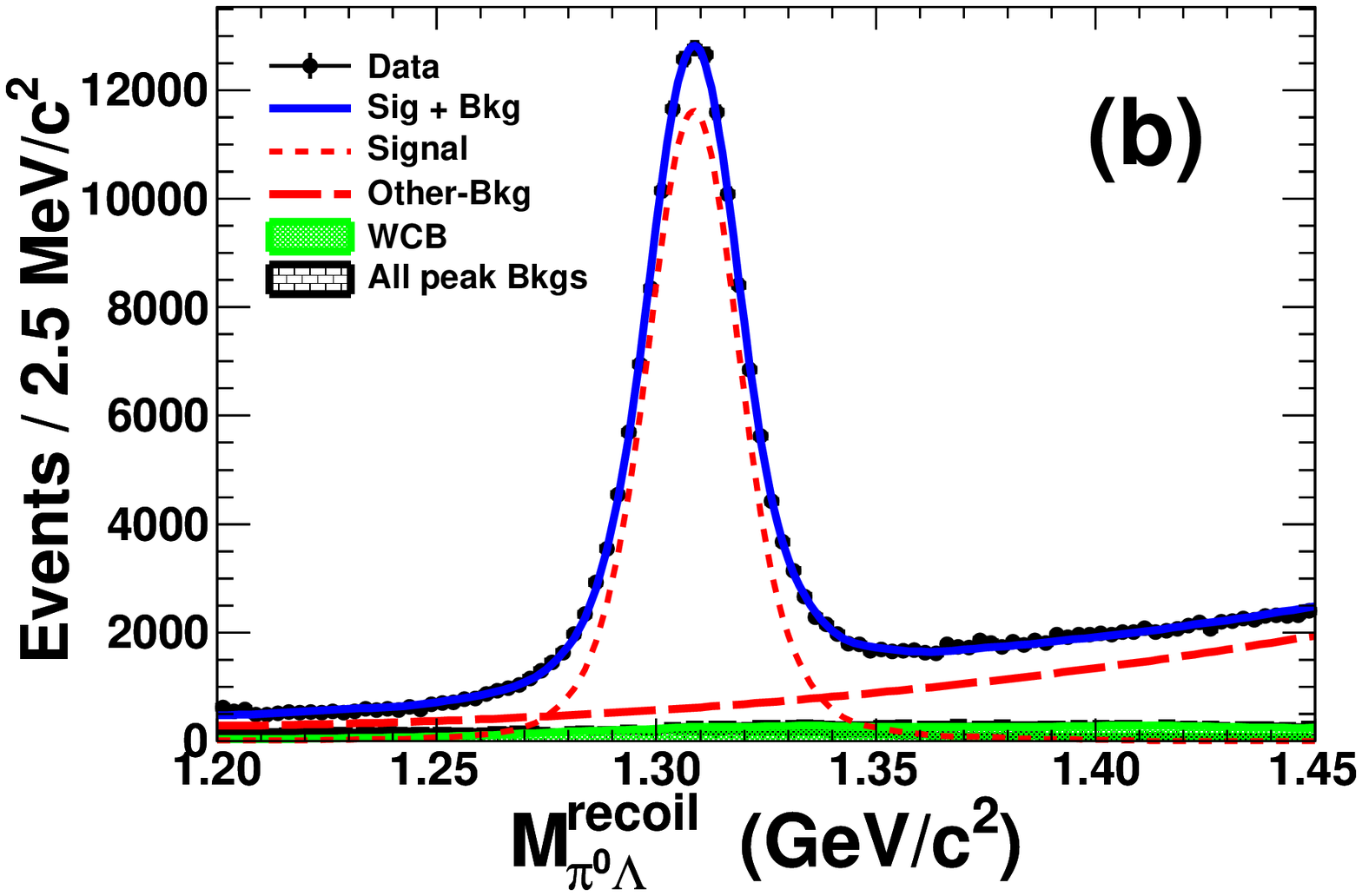}}\\
\subfigure{\includegraphics[width=0.45\textwidth]{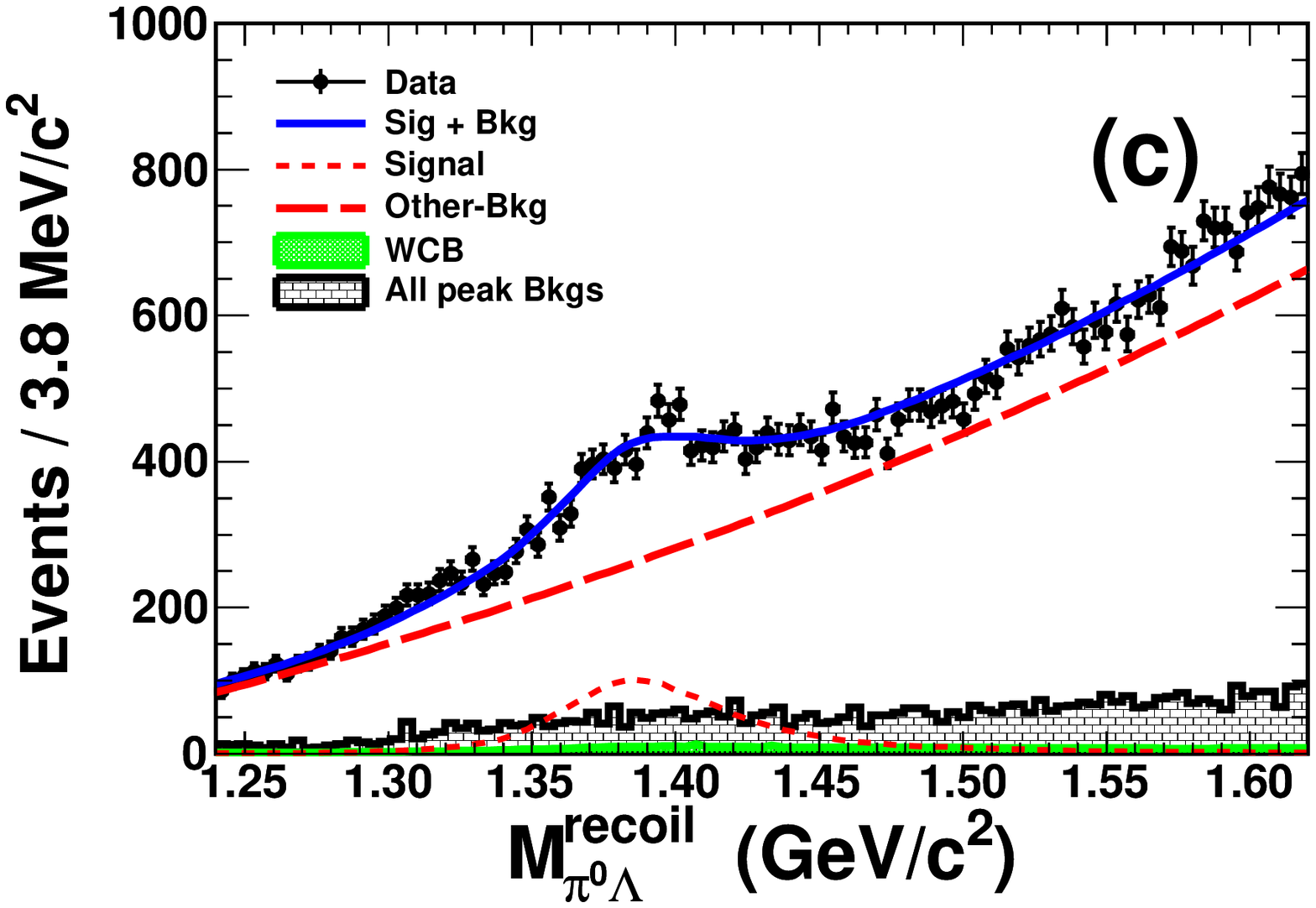}}
\subfigure{\includegraphics[width=0.45\textwidth]{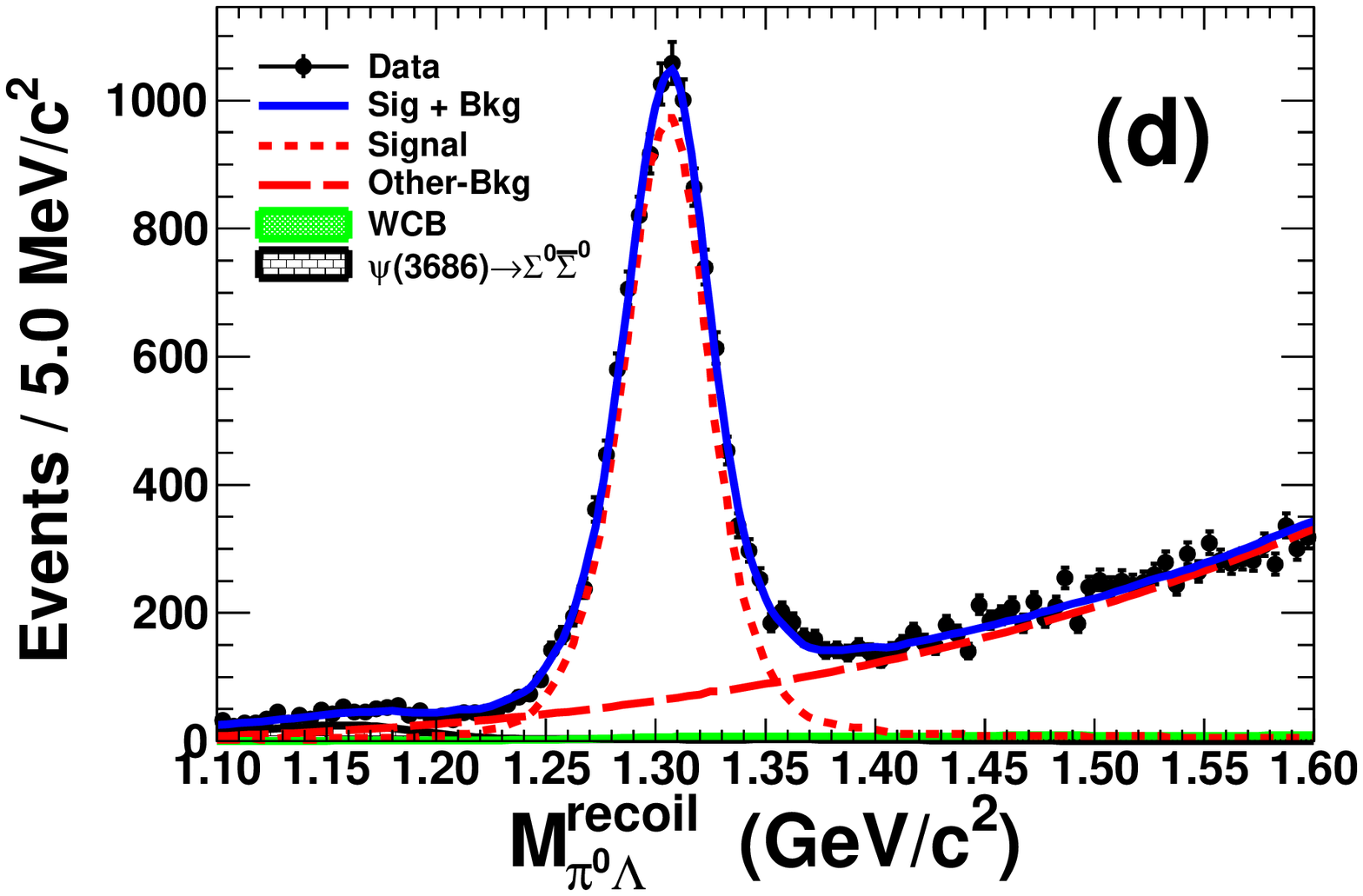}}
\caption{(Color online). Recoil mass spectra of $\pi^{0}\Lambda$ for (a) $\jpsi\ar\SSSN$, (b) $\jpsi\ar\XXN$, (c) $\psp\ar\SSSN$, and (d)  $\psp\ar\XXN$. Dots with error bars indicate the data, the blue solid lines show the fit result, the red short-dashed lines are for signal, the red long-dashed ones are for the remaining background (Other-Bkg), and the green hatched ones are for wrong combination background (WCB) , the black hatched ones are for the peaking backgrounds.}
\label{fitting}
\ecl
\end{figure*}

The signal yields for the decays $\psi\ar\SSSN$ and $\XXN$ are extracted by performing an extended maximum likelihood fit to the $M^\text{recoil}_{\pi^{0}\Lambda}$ spectrum. In the fit, the signal shape is represented by the simulated MC shape convolved with a Gaussian function to take into account the mass resolution difference between data and MC simulation.
The peaking backgrounds and the wrong combination background are described by the individual shape taken from MC simulation, and the corresponding numbers of background events are fixed according to the individual detection efficiencies and branching fractions~\cite{PDG2012}. The remaining backgrounds are found to be distributed smoothly  in the $M_{\pi^0\Lambda}^{\rm recoil}$ spectrum and are therefore described by a second-order polynomial function.
Figure~\ref{fitting} shows the projection plots of $M^{\rm recoil}_{\pi^{0}\Lambda}$ for the decays $\psi\ar\SSSN$ and $\XXN$, respectively.

The branching fraction can be calculated by
\begin{center}
${\cal B}[\psi\ar X\bar{X}]=\frac{N_{\rm obs}}{N_{\psi}\cdot\epsilon \cdot{\cal B}(X\ar\pi^0\Lambda) \cdot {\cal B}(\Lambda\ar p\pi) \cdot {\cal B}(\pi^{0}\ar\gamma\gamma)}$,
\end{center}
where
$X$ stands for the $\Sigma(1385)^{0}$ or $\Xi^{0}$ baryon, 
$\epsilon$ denotes the detection efficiency obtained with the measured $\alpha$ value, $N_{\rm obs}$ is the number of observed signal events, ${\cal B}(X\ar\pi^0\Lambda)$, ${\cal B}(\Lambda\ar p\pi)$ and ${\cal B}(\pi^{0}\ar\gamma\gamma)$ are the branching fractions of $X\ar\Lambda\pi^0$, $\Lambda\ar p\pi$ and $\pi^{0}\ar\gamma\gamma$ taken from PDG~\cite{PDG2012}, $N_{\psi}$ is the total number of $\jpsi$ or $\psp$ events~\cite{Jpsi, Psip}. Table~\ref{result} summarizes the numbers of observed signal events, the corresponding efficiencies, and branching fractions for the various decays in this measurement with the statistic uncertainty only.
\subsection{Angular distribution}
The values of $\alpha$ for the four decay processes are determined by performing a least squares fit to the $\cos\theta$ distribution in the range from $-0.8$ to $0.8$, divided into 8 equidistant intervals for the decays $\psp\ar\SSSN$ and into 16 intervals for the other three decay modes.

The signal yield in each $\cos\theta$ bin is obtained with the
aforementioned fit method.
The distributions of the efficiency-corrected signal yields together with the fit curves are
shown in Fig.~\ref{angular}.
The $\alpha$ values obtained from the fits based on Eq.~(\ref{alpha}) are summarized in Table~\ref{result}.

\btbl[!htb]
 \caption{The numbers of the observed events $N_{\rm obs}$, efficiencies $\epsilon$, $\alpha$ values, and branching fractions ${\cal B}$ for $\psi\ar\SSSN$ and $\XXN$. Only the statistical uncertainties are indicated.}
  \bcl
  \newcommand{\ST}{\rule[0ex]{0pt}{3ex}}
\begin{tabular}{lcccc}  \hline 
Channel            &$N_{\rm obs}$              &$\epsilon$(\%)     &$\alpha$            &${\cal B} (\times 10^{-4}$)  \ST \\ \hline
$\jpsi\ar\SSSN$    &102762 $\pm$ 852   &13.32 $\pm$ 0.04   &$-0.64 \pm 0.03$   &$~$10.71 $\pm$ 0.09  \ST \\
$\jpsi\ar\XXN$     &134846 $\pm$ 437   &14.05 $\pm$ 0.04   &$~~~$0.66 $\pm$ 0.03     &$11.65 \pm 0.04$  \\
$\psp\ar\SSSN$     &$~~~~$2214  $\pm$  149 &13.13 $\pm$ 0.03   &$~~~$0.59 $\pm$ 0.25  &$~~$0.69 $\pm$ 0.05 \\
$\psp\ar\XXN$      &$~~$10839 $\pm$ 123    &14.10 $\pm$ 0.04   &$~~~$0.65 $\pm$ 0.09  &$~~$2.73 $\pm$ 0.03 \\  \hline 
\end{tabular}
\label{result}
\ecl
\etbl
\section{Systematic uncertainty}
\label{sec:syst_err}
\subsection{Branching fraction}
Systematic uncertainties on the branching fractions are mainly due to efficiency differences
between data and MC simulation.  They are
estimated by comparing the efficiencies of photon, $\pi^{0}$, $\Lambda$ and $\Xi^{0}$ reconstruction between the data and the MC simulation. Additional sources of systematic uncertainties are the
fit range, wrong combination, the background shape, and the angular distributions. In addition, the uncertainties of the decay branching fractions of intermediate states and uncertainties of the total number of $\psi$ events are also accounted for in the systematic uncertainty.
All of the systematic uncertainties are discussed in detail below.

\begin{enumerate}
    \item The uncertainty associated with photon detection efficiency is 1.0\% per photon, which is determined using the control sample  $\jpsi\ar\rho\pi$. Hence, for $\psi\ar\SSSN$, the value 2.0\% is taken as the systematic uncertainty.

    \item The systematic uncertainty due to the 1C kinematic fit for the $\pi^{0}$ reconstruction is estimated to be 1.0\% with the control sample $\jpsi\ar\rho\pi$.
    \item The uncertainty related to the $\Lambda$ reconstruction efficiency in $\Sigma(1385)$ decays is estimated using the control sample $\psi\ar\XXB$. Here, the $\Lambda$ reconstruction efficiency includes systematic uncertainties due to tracking, PID, and the vertex fit. A detailed description of this method can be found
in Ref.~\cite{wangxf}. 
\begin{figurehere}
\bcl
\subfigure{\includegraphics[width=0.35\textwidth]{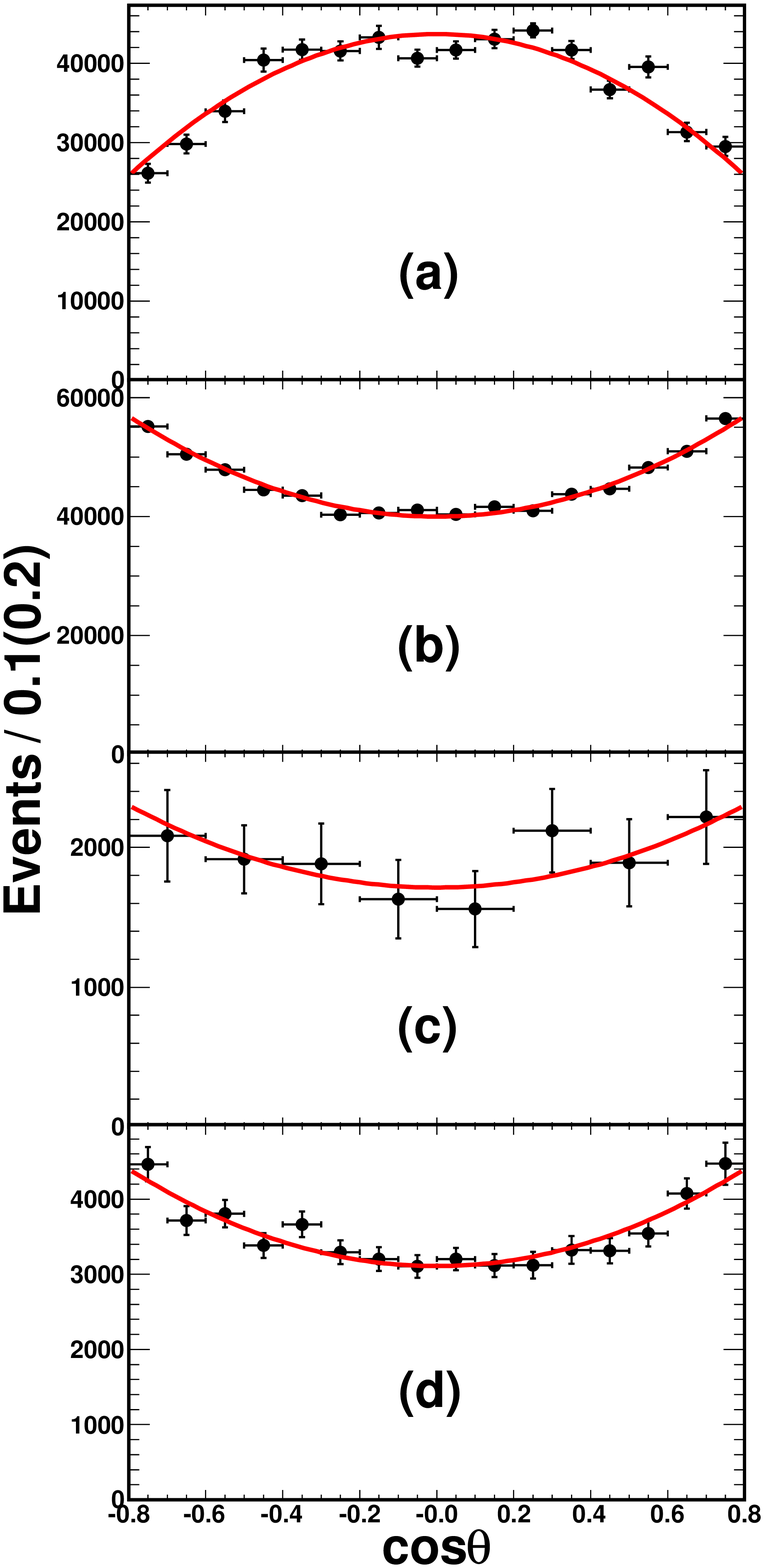}}
\caption{Distributions of $\cos\theta$ for (a) $\jpsi\ar\SSSN$,(b) $\jpsi\ar\XXN$, (c) $\psp\ar\SSSN$, and (d) $\psp\ar\XXN$. The dots with error bars indicate the efficiency corrected data, and the curves show the fit results.}
\label{angular}
\ecl
\end{figurehere}

   \item The $\Xi^{0}$ reconstruction efficiency, which includes the two photon efficiencies, $\pi^{0}$ reconstruction efficiency and the $\Lambda$ reconstruction efficiency, is studied with the control sample $\jpsi\ar\XXN$ 
via single and double tag methods. The selection criteria of the charged tracks, and the reconstruction of $\Lambda$ and $\Xi^{0}$ candidates are exactly same as those described in Sec.~\ref{sec:evt_sel}. The $\Xi^{0}$ reconstruction efficiency is defined as the ratio of the number of events from the double tag $\XXN$ to that from the single tag. The difference in the $\Xi^{0}$ reconstruction efficiency between data and MC samples is taken as the systematic uncertainty.

    \item In the fits of the $M^{\rm recoil}_{\pi^{0}\Lambda}$ signal, the uncertainty due to the fitting range is estimated by varying the mass range by $\pm$ 10 MeV/$c^{2}$ for two sides. The resulting differences of signal yields are taken as the systematic uncertainty.
     \item The uncertainties due to the background shape arise from the polynomial function and the peaking shape. The former is estimated by the alternative fits with a first or a third-order polynomial function.
         The latter is estimated by varying the number of normalized events by 1$\sigma$. The larger difference is taken as the systematic uncertainty.
         The total uncertainty related to the background shape is obtained by adding all contributions in quadrature.
     \item The systematic uncertainty due to the wrong combination background is estimated by comparing the signal yields between the fits  with and without the corresponding component included in the fit. The differences of signal yields are taken as systematic uncertainties.
    \item The uncertainty related with the detection efficiency due to the modeling of the angular distribution of the baryon pairs,
         represented by the parameter $\alpha$, is estimated by varying the measured $\alpha$ values by 1$\sigma$ in the MC simulation. The changes in the detection efficiency are taken as a systematic uncertainty.
    \item The systematic uncertainties due to the branching fractions of the intermediate states, $\Xi^{0}$, $\Sigma(1385)^{0}$ and $\Lambda$, are taken from the PDG~\cite{PDG2012}.  They are 1.9\% for $\psi\ar\SSSN$ and 0.8\% for $\psi\ar\XXN$.
    \item The systematic uncertainties due to the total number of $\jpsi$ or $\psp$ events are determined with the inclusive hadronic $\psi$ decays.  They are 0.5\% and 0.6\% in ~\cite{Jpsi,Psip}, respectively.
\end{enumerate}

The various systematic uncertainties on the branching fraction measurements
 are summarized in Table~\ref{error}. The
total systematic uncertainty is obtained by summing the individual
contributions in quadrature.

\btbl[!htb]
\caption{Relative systematic uncertainties on the branching fraction measurements (in \%).}
\bcl
\doublerulesep 2pt
\begin{tabular}{ccccc}  \hline
\multicolumn{1}{c}{}   &\multicolumn{2}{c}{$\jpsi\ar$} &\multicolumn{2}{c}{$\psp\ar$} \\ \hline
Source                          &$\SSSN$  &$\XXN$  &$\SSSN$ &$\XXN$ \\ \hline
Photon efficiency             &2.0      &...      &2.0     &...\\
$\pi^{0}$ reconstruction      &1.0      &...     &1.0      &... \\
$\Lambda$ reconstruction      &3.0      &...     &1.0      &...\\
$\Xi^{0}$ reconstruction      &...      &2.6     &...      &2.6\\
Fit range                     &2.1      &1.6     &2.8     &1.8\\
Background shape              &3.9      &1.5     &4.0     &2.3\\
Wrong combination             &4.2      &0.8     &4.5     &0.3 \\
Angular distribution          &2.0      &0.5     &1.2     &2.8\\
Intermediate decay            &1.9      &0.8     &1.9     &0.8 \\
Total number of $\psi$        &0.5      &0.5     &0.6     &0.6 \\ 
Total                         &7.7      &3.7     &7.4    &4.9\\ \hline 
\end{tabular}
\label{error}
\ecl
\etbl

\subsection{Angular distribution}
Various systematic uncertainties are considered in the measurement of the values of $\alpha$. These include the uncertainty of the
signal yield in the different $\cos\theta$ intervals, the uncertainty of the
$\cos\theta$ fit procedure, and the uncertainty related to the detection
efficiency correction curve as function of the $\cos\theta$ bin. They are discussed in detail below.
\begin{enumerate}
\item The signal yields in each $\cos\theta$ interval are determined by the
    fit to the corresponding $M^{\rm recoil}_{\pi^{0}\Lambda}$ distribution.
    The sources of the systematic uncertainty of the signal yield include
    the fit range, the background shape, MC resolution and wrong combination, where the MC resolution is fixed for the decay $\psp\ar\SSSN$ only. To estimate the systematic uncertainty related with fit range on $M^{\rm recoil}_{\pi^{0}\Lambda}$, we repeat the fit to the $M^{\rm recoil}_{\pi^{0}\Lambda}$ distribution by changing the fit range
by $\pm$10 MeV/$c^{2}$.
Then, the $\alpha$ values are extracted by the fit with the changed signal yield,
and the resulting differences to the nominal $\alpha$ values are taken
as the systematic uncertainties. The uncertainties related to the background shape, MC resolution and wrong combination backgrounds in the fit are evaluated with a method similar to the one described above.
\item The systematic uncertainties related to the procedure of the fit on the $\cos\theta$ distribution are estimated by
re-fitting the $\cos\theta$ distribution with a different binning and fit range. We divide $\cos\theta$ into 8 intervals
for $\psi\ar\XXN$ and 16 intervals for $\psi\ar\SSSN$. The changes of the $\alpha$ values are taken as systematic
uncertainties. We also repeat the fit by changing the range to [$-$0.9, 0.9] or [$-$0.7, 0.7] in $\cos\theta$, with the same
 bin size of the nominal fit. The largest differences of $\alpha$ value with respect to the nominal value are taken as a systematic uncertainties.

\item 
In the analysis, the $\alpha$ values are obtained by fitting the $\cos\theta$ distribution corrected by the detection efficiency. To estimate the systematic uncertainty related to the imperfect simulation of the detection efficiency, the ratio of detection efficiencies as function of $\cos\theta$ between data and MC simulation is obtained based on the control sample $\jpsi\ar\XXN$ with a full event reconstruction. Then, the efficiency corrected $\cos\theta$ distribution scaled by the ratios of detection efficiencies is refitted.  The resulting differences in $\alpha$ are taken as the systematic uncertainty.
\end{enumerate}

All the systematic uncertainties for the $\alpha$ measurement are summarized in Table~\ref{error_ang}. The total systematic
uncertainty is
the quadratic sum of the individual values.

\btbl[!htb]
\caption{Relative systematic uncertainties on the $\alpha$ value measurements (in \%).}
\bcl
\doublerulesep 2pt
\begin{tabular}{ccccc}  \hline
\multicolumn{1}{c}{} &\multicolumn{2}{c}{$\jpsi\ar$} &\multicolumn{2}{c}{$\psp\ar$} \\ \hline
Source                                   &$\SSSN$ &$\XXN$  &$\SSSN$ &$\XXN$ \\ \hline
$M^{\rm recoil}_{\pi^{0}\Lambda}$ fitting range &7.8    &3.0      &15.3   &7.7\\
Background shape                            &3.2    &3.0      &20.0   &4.6\\
MC resolution                                    &...     &...       &16.9   &...\\
Wrong combination                           &4.7   &1.5      &5.1    &15.0\\
$\cos\theta$ interval                         &7.8    &3.5      &22.0   &10.4 \\
$\cos\theta$ fitting range                  &7.8    &3.0      &15.6   &3.5\\
Efficiency correction                        &4.7    &3.0      &9.0    &3.0\\ 
Total                                                &15.4   &7.1      &41.8   &20.8\\ \hline 
\end{tabular}
\label{error_ang}
\ecl
\etbl
\section{Conclusion and discussion}
\label{sec:conclusion}
Using $(1310.6 \pm 7.0) \times 10^{6}$ $\jpsi$ and $(447.9 \pm 2.9) \times 10^{6}$ $\psp$ events
collected with the BESIII detector at BEPCII, the branching fractions and the angular distributions
for $\psi\ar\SSSN$ and $\XXN$ are measured.
A comparison of the branching fractions  between our measurement and previous experiments (PDG average) is summarized in Table~\ref{result02}.  The branching fractions for $\psi\ar\SSSN$ are measured for the first time, 
and the branching fractions for $\psi\ar\XXN$ are measured with a good agreement and a much higher precision than the previous  results.
The measured $\alpha$ values are also compared with the predictions of the theoretical models from Refs.~\cite{ppbref02, ppbref01}.
As indicated in Table~\ref{result01}, some of our results disagree significantly with the theoretical predictions, which may imply that the naive prediction of QCD suffers from the approximation that higher-order corrections are not taken into account.
As calculated in Ref.~\cite{Chen:2006yn}, the sign for parameter $\alpha$ in $\psi\ar\ssb$ mode could be negative if re-scattering effects in the final states are taken into account. However, our results show that $\alpha$ for $\jpsi$ 
is negative, and is different to the other decay processes in this measurement,
 which is hard to explain within the existing models.
We, therefore, believe that it is of utmost importance to improve the theoretical models to shed further light on the origin of these discrepancies.

\btbl[!htb]
\caption{
Comparison of the branching fractions for $\psi\ar\SSSN$ and $\XXN$ (in units of $ 10^{-4}$). The first uncertainties are statistical, and the second systematic.}
\bcl
\footnotesize{
\doublerulesep 2pt
\begin{tabular}{ccccc}  \hline
Mode            &$\jpsi\ar\SSSN$ &$\jpsi\ar\XXN$ &$\psp\ar\SSSN$  &$\psp\ar\XXN$\\ \hline
This work          &$10.71 \pm 0.09 \pm 0.82$  &$11.65 \pm 0.04 \pm 0.43$ &$0.69 \pm 0.05 \pm 0.05$ &$2.73 \pm 0.03 \pm 0.13$\\
BESII~\cite{Ablikim:2008tj}   &...&$12.0 \pm 1.2 \pm 2.1$&... &...\\
CLEO~\cite{Pedlar:2005px}&...&...&... &$2.75 \pm 0.64 \pm 0.61$\\
Dobbs~\emph{et al.}~\cite{Dobbs:2014ifa}&...&...&... &$2.02 \pm 0.19 \pm 0.15$\\
PDG~\cite{PDG2012}            &...&$12.0 \pm 2.4$&... &$2.07 \pm 0.23$\\  \hline 
\end{tabular}}
\label{result02}
\ecl
\etbl

\btbl[!htb]
\caption{Comparison of the $\alpha$ values for $\psi\ar\SSSN$ and $\XXN$, the first uncertainties are statistical
and the second systematic.}
\bcl
\footnotesize{
\doublerulesep 2pt
\begin{tabular}{ccccc}  \hline 
Mode                           &$\jpsi\ar\SSSN$         &$\jpsi\ar\XXN$
                               &$\psp\ar\SSSN$          &$\psp\ar\XXN$ \\ \hline
This work                      &$-$0.64 $\pm$ 0.03 $\pm$ 0.10 &$0.66 \pm 0.03 \pm 0.05$
                               &0.59 $\pm$ 0.25 $\pm$ 0.25  &$0.65 \pm 0.09 \pm 0.14$  \\
Carimalo \emph{et al.}~\cite{ppbref02}   &0.11      &0.16  &0.28 &0.33          \\
Claudson~\cite{ppbref01}   &0.19      &0.28  &0.46 &0.53         \\ \hline 
\end{tabular}}
\label{result01}
\ecl
\etbl
To test the \textquotedblleft 12\% rule", the ratios of the  branching fractions  $\frac{{\cal{B}}(\psp\ar\SSSN)}{{\cal{B}}(\jpsi\ar\SSSN)}$ and $\frac{{\cal{B}}(\psp\ar\XXN)}{{\cal{B}}(\jpsi\ar\XXN)}$  are calculated to be (6.44 $\pm$ 0.47 $\pm$ 0.64)\% and (23.43 $\pm$ 0.26 $\pm$ 1.09)\%, respectively, taking into account the cancelation of the common systematic uncertainties.
The ratios are not in agreement with 12\%, especially for the $\XXN$ final state.

To test isospin symmetry, the ratios of the branching fractions  listed in Table~\ref{QIS} are also calculated based on the measurements between the neutral mode and the corresponding charged modes~\cite{Ablikim:2016iym} taking into account the cancelation of the common systematic uncertainties. All ratios are within $1\sigma$ of the expectation of isospin symmetry.

\btbl[!htb]
\caption{
Summary of the ratios of branching fraction for testing isospin symmetry. The first uncertainties are the statistical, and the second systematic.}
\bcl
\footnotesize{
\doublerulesep 2pt
\begin{tabular}{ccccc}  \hline
Mode            &$\frac{{\cal{B}}(\psi\ar\XXN)}{{\cal{B}}(\psi\ar\XXB)}$
                    &$\frac{{\cal{B}}(\psi\ar\SSSN)}{{\cal{B}}(\psi\ar\SSSM)}$
                    &$\frac{{\cal{B}}(\psi\ar\SSSN)}{{\cal{B}}(\psi\ar\SSSP)}$ \\ \hline
$\jpsi$          &$1.12 \pm 0.01 \pm 0.07$ &$0.98 \pm 0.01 \pm 0.08$ &$0.85 \pm 0.02 \pm 0.09$\\  
$\psp$         &$0.98 \pm 0.02 \pm 0.07$ &$0.81 \pm 0.12 \pm 0.12$ &$0.82 \pm 0.11 \pm 0.11$ \\ \hline 
\end{tabular}}
\label{QIS}
\ecl
\etbl
\section{Acknowledgement}
\label{sec:acknowledgement}
The BESIII collaboration thanks the staff of BEPCII and the IHEP computing center for their strong support. This work is supported in part by National Key Basic Research Program of China under Contract No. 2015CB856700; National Natural Science Foundation of China (NSFC) under Contracts Nos. 11235011, 11322544, 11335008, 11425524, 11475207, 11505034, 11565006, 11635010; the Chinese Academy of Sciences (CAS) Large-Scale Scientific Facility Program; the CAS Center for Excellence in Particle Physics (CCEPP); the Collaborative Innovation Center for Particles and Interactions (CICPI); Joint Large-Scale Scientific Facility Funds of the NSFC and CAS under Contracts Nos. U1232107, U1232201, U1332201, U1532257, U1532258; CAS under Contracts Nos. KJCX2-YW-N29, KJCX2-YW-N45; 100 Talents Program of CAS; National 1000 Talents Program of China; INPAC and Shanghai Key Laboratory for Particle Physics and Cosmology; German Research Foundation DFG under Contracts Nos. Collaborative Research Center CRC 1044, FOR 2359; Istituto Nazionale di Fisica Nucleare, Italy; Koninklijke Nederlandse Akademie van Wetenschappen (KNAW) under Contract No. 530-4CDP03; Ministry of Development of Turkey under Contract No. DPT2006K-120470; The Swedish Resarch Council; U. S. Department of Energy under Contracts Nos. DE-FG02-05ER41374, DE-SC-0010504, DE-SC-0010118, DE-SC-0012069; U.S. National Science Foundation; University of Groningen (RuG) and the Helmholtzzentrum fuer Schwerionenforschung GmbH (GSI), Darmstadt; WCU Program of National Research Foundation of Korea under Contract No. R32-2008-000-10155-0.

\end{multicols}
\end{document}